\documentclass[11pt]{cernrep}
\usepackage{graphicx}
\usepackage{here}

\begin{document}
\bibliographystyle{unsrt}
\renewcommand{\thefootnote}{\fnsymbol{footnote}}

\begin{flushright}
hep-ph/0401145\\
LU TP 04--07\\
January 2004
\end{flushright}

\vspace{10mm}

\noindent{\Large\bf RESUMMATION AND SHOWER STUDIES%
\footnote{submitted to the proceedings of the Workshop on 
Physics at TeV Colliders, Les Houches, France, 
26 May -- 6 June 2003}}\\[8mm]
\textit{J. Huston$^1$, I. Puljak$^2$, T. Sj\"ostrand$^3$, E. Thom\'e$^3$}\\
$^1$Department of Physics and Astronomy, Michigan State University, USA\\ 
$^2$FESB, University of Split, Split, Croatia\\
$^3$Department of Theoretical Physics, Lund University, Sweden

\section{INTRODUCTION}

The transverse momentum of a colour-singlet massive particle produced 
in a hadronic collision provides important information on perturbative 
and nonperturbative effects. A process like 
$\mathrm{q} \overline{\mathrm{q}} \to \mathrm{Z}^0$ corresponds to a 
$p_{\perp\mathrm{Z}} = 0$, while higher-order processes 
provide $p_{\perp}$ kicks as the $\mathrm{Z}^0$ recoils against quarks
and gluons. At large $p_{\perp\mathrm{Z}}$ values the bulk of the
$p_{\perp}$ comes from one hard emission, and perturbation theory is a 
reasonable approach. In the small-$p_{\perp\mathrm{Z}}$ region, on the 
other hand, many emissions can contribute with $p_{\perp}$ kicks of 
comparable size, and so the order-by-order approach is rather poorly 
convergent. Furthermore, in this region nonperturbative effects may 
start to become non-negligible relative to the perturbative ones.

The traditional solution has been to apply either an analytical 
resummation approach or a numerical parton-shower one. These methods 
to some extent are complementary. The norm today is for showers to be
based on an improved leading-log picture, while resummation is carried 
out to next-to-leading logs. 
However, resummation gives no information on the partonic system 
recoiling against the $\mathrm{Z}^0$, while showers do, and therefore 
can be integrated into full-fledged event generators, allowing accurate 
experimental studies. In both approaches the high-$p_{\perp}$ tail is 
constrained by fixed-order perturbation theory, so the interesting and 
nontrivial region is the low-to-medium-$p_{\perp}$ one. Both also require 
nonperturbative input to handle the low-$p_{\perp}$ region, e.g.\ in 
the form of a primordial $k_{\perp}$ carried by the initiator of a 
shower. 

One of the disconcerting aspects of the game is that a large primordial 
$k_{\perp}$ seems to be required and that the required value of this
primordial $k_{\perp}$ can be dependent on the kinematics of the process 
being considered. Confinement of partons inside the 
proton implies a $\langle k_{\perp} \rangle \approx 0.3$~GeV, while 
fits to $\mathrm{Z}^0$ data  at the Tevatron favour $\approx 2$~GeV 
\cite{Balazs:2000sz} (actually as a root-mean-square value, assuming 
a Gaussian distribution). Also resummation approaches tend to require 
a non-negligible nonperturbative contribution,
but that contribution can be determined from fixed-target data and then
automatically evolved to the kinematical region of interest. 
In this note we present updated comparisons and study 
possible shower modifications that might alleviate the problem. We will 
use the two cases of $\mathrm{q} \overline{\mathrm{q}} \to \mathrm{Z}^0$ 
and $\mathrm{g} \mathrm{g} \to \mathrm{H}^0$ (in the infinite-top-mass
limit) to illustrate differences in quark and gluon evolution, and
the Tevatron and the LHC to quantify an energy dependence.

\section{COMPARISON STATUS}

A detailed comparison of analytic resummation and parton showers 
was presented in \cite{Balazs:2000sz}. For many physical quantities, 
the predictions from parton shower Monte Carlo programs should be nearly
as precise as those from analytical theoretical calculations. In 
particular, both analytic and parton shower Monte Carlos should accurately 
describe the effects of the emission of multiple soft gluons from the 
incoming partons. 

Parton showers resum primarily the leading logs, which are universal, i.e.
process-independent, depending only on the initial state. An analytic 
resummmation calculation, in principle, can resum all logs, but in practice
the number of towers of logarithms included in the analytic Sudakov 
exponent depends on the level to which a fixed-order calculation was 
performed for a given process. Generally, if a NNLO calculation is 
available, then the $B^{(2)}$ coefficient (using the CSS formalism
\cite{Collins:1981uk}) can be extracted and incorporated. If we try 
to interpret parton showering in the same language then we say that 
the Monte Carlo Sudakov exponent always contains a term analogous to 
$A^{(1)}$ and $B^{(1)}$ and that an approximation to $A^{(2)}$ is also 
present in some kinematic regions. 

In Ref.~ \cite{Balazs:2000sz}, predictions were made for both 
$\mathrm{Z}^0$ and Higgs production at the Tevatron and the  LHC, 
using both resummation and parton shower Monte Carlo programs. In 
general, the shapes for the $p_{\perp}$ distributions agreed well, 
although the \textsc{Pythia} showering algorithm typically caused 
the Higgs $p_{\perp}$ distribution to peak at somewhat lower values 
of transverse momentum. 

\section{SHOWER ALGORITHM CONSTRAINTS}

While customarily classified as leading log, shower algorithms
tend to contain elements that go beyond the conventional leading-log
definition. Specifically, some emissions allowed by leading log
are forbidden in the shower description. Taking the \textsc{Pythia}
\cite{Sjostrand:2000wi,Sjostrand:2003wg} initial-state shower algorithm 
\cite{Sjostrand:1985xi,Bengtsson:1986gz,Miu:1998ju} as an example,
the following aspects may be noted (see \cite{Thome:2004sk} for
further details):\\
\textit{(i)} Emissions are required to be angularly ordered, such that 
opening angles increase on the way in to the hard scattering subprocess.
That is, non-angularly-ordered emissions are vetoed.\\
\textit{(ii)} The $z$ and $Q^2$ of a branching $a \to b c$ are required 
to fulfill the condition $\hat{u} = Q^2 - \hat{s}(1-z) < 0$. Here 
$\hat{s} = (p_a + p_d)^2 = (p_b + p_d)^2/z$, for $d$ the incoming parton 
on the other side of the event. In the case that $b$ and $d$ form a 
$\mathrm{Z}^0$, say, and $c$ is the recoiling parton, $\hat{u}$ 
coincides with the standard Mandelstam variable for the 
$a + d \to (\mathrm{Z}^0=b+d) + c$ process. In general, it may be 
viewed as a kinematics consistency constraint.\\
\textit{(iii)} The evolution rate is proportional to
$\alpha_{\mathrm{s}}((1-z) Q^2) \approx 
\alpha_{\mathrm{s}}(p_{\perp}^2)$ rather than 
$\alpha_{\mathrm{s}}(Q^2)$. Since $p_{\perp}^2 < Q^2$ by itself this 
implies a larger $\alpha_{\mathrm{s}}$ and thus an increased rate of 
evolution. However, one function of the $Q_0 \approx 1$~GeV 
nonperturbative cutoff parameter is to avoid the 
divergent-$\alpha_{\mathrm{s}}$ region, so now one must require
$(1-z) Q^2 > Q_0^2$ rather than $Q^2 > Q_0^2$. The net result again
is a reduced emission rate.\\
\textit{(iv)} One of the partons of a branching may develop a timelike
parton shower. The more off-shell this parton, the less the $p_{\perp}$
of the branching. The evolution rate in $x$ is unaffected, however.\\
\textit{(v)} There are some further corrections, that in practice appear
to have negligible influence: the non-generation of very soft gluons to 
avoid the divergence of the splitting kernel, the possibility of photon
emission off quarks, and extra kinematical constraints when heavy quarks 
are produced.\\
\textit{(vi)} The emission rate is smoothly merged with the first-order 
matrix elements at large $p_{\perp}$. This is somewhat separate from
the other issues studied, and the resulting change only appreciably 
affects a small fraction of the total cross section, so it will not be 
considered further here.

The main consequence of the first three points is a lower rate of $x$
evolution. That is, starting from a set of parton densities 
$f_i(x, Q_0^2)$ at some low $Q_0^2$ scale, and a matching $\Lambda$, 
tuned such that standard DGLAP evolution provides a reasonable fit to 
data at $Q^2 > Q_0^2$, the constraints above lead to $x$ distributions 
less evolved and thus harder than data. If we e.g.\ take the CTEQ5L tune 
\cite{Lai:1999wy} with $\Lambda^{(4)} = 0.192$~GeV, the $\Lambda^{(4)}$ 
would need to be raised to about 0.23~GeV in the shower to give the same 
fit to data as CTEQ5L when the angular-ordering cut in \textit{(i)} is 
imposed. Unfortunately effects from points \textit{(ii)} and  
\textit{(iii)} turn out to be process-dependent, presumably reflecting 
kinematical differences between $\mathrm{q}\to\mathrm{q}\mathrm{g}$ and
$\mathrm{g} \to \mathrm{g}\mathrm{g}$. There is also some energy 
dependence. The net result of the first three points suggests that 
\textsc{Pythia} should be run with a $\Lambda^{(4)}$ of about 0.3~GeV
(0.5~GeV) for $\mathrm{Z}^0$ ($\mathrm{H}^0$) production in order to 
compensate for the restrictions on allowed branchings.

One would expect the increased perturbative evolution to allow the 
primordial $k_{\perp}$ to be reduced. Unfortunately, while the total
radiated transverse energy, $\sum_i |\mathbf{p}_{\perp i}|$, comes up 
by about 10\% at the Tevatron, this partly cancels in the vector sum, 
$\mathbf{p}_{\perp\mathrm{Z}} = - \sum_i \mathbf{p}_{\perp i}$. For a
2~GeV primordial $k_{\perp}$ the shift of the peak position of the 
$p_{\perp\mathrm{Z}}$ spectrum is negligible. Results are more visible 
for $p_{\perp\mathrm{H}}$ at the LHC.

Note that a primordial $k_{\perp}$ assigned to the initial parton at
the low $Q_0^2$ scale is shared between the partons at each shower 
branching, in proportion to the longitudinal momentum fractions a 
daughter takes. Only a fraction $x_{\mathrm{hard}}/x_{\mathrm{initial}}$
of the initial $k_{\perp}$ thus survives to the hard-scattering parton.
Since the typical $x$ evolution range is much larger at the LHC than at
the Tevatron, a tuning of the primordial $k_{\perp}$ is hardly an option
for $\mathrm{H}^0$ at the LHC, while it is relevant for $\mathrm{Z}^0$ at 
the Tevatron. Therefore an increased $\Lambda$ value is an interesting 
option. 
 
We now turn to the point \textit{(iv)} above. By coherence 
arguments, the main chain of spacelike branchings sets the maximum
virtuality for the emitted timelike partons, i.e.\ the timelike 
branchings occur at longer timescales than the related spacelike ones. 
In a dipole-motivated language, one could therefore imagine that the 
recoil, when a parton acquires a timelike mass, is not taken by a 
spacelike parton but by other final-state colour-connected partons. A 
colour-singlet particle, like the $\mathrm{Z}^0$ or $\mathrm{H}^0$, 
would then be unaffected by the timelike showers. 
    
The consequences for $p_{\perp\mathrm{Z}}$ and $p_{\perp\mathrm{H}}$
of such a point of view can be studied by switching off timelike showers 
in \textsc{Pythia}, but there is then no possibility to fully simulate 
the recoiling event. A new set of shower routines is in preparation 
\cite{Sjostrand:2004qj}, however, based on $p_{\perp}$-ordered emissions 
in a hybrid between conventional showers and the dipole approach. It is 
well suited for allowing final-state radiation at later times, leaving 
$p_{\perp\mathrm{Z}}$ and $p_{\perp\mathrm{H}}$ unaffected at that 
stage. Actually, without final-state radiation, the two approaches give 
surprisingly similar results overall. Both are lower in the peak region 
than the algorithm with final-state radiation, in better agreement with 
CDF data \cite{Affolder:1999jh}. The new one is slightly lower, i.e.\ 
better relative to data, at small $p_{\perp\mathrm{Z}}$ values. 

\begin{figure}[pt]
\begin{center}
\vspace*{-5mm}
\includegraphics[width=12cm]{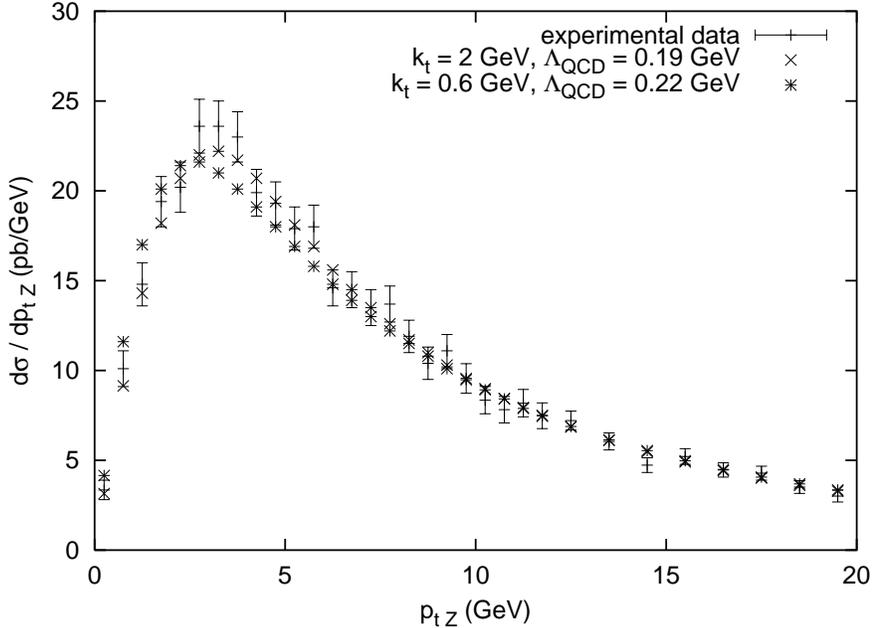}
\vspace*{-5mm}
\end{center}
\caption{Comparison of the CDF $p_{\perp\mathrm{Z}}$ spectrum with the 
new shower algorithm for two parameter sets.}
\label{resumandshower:figa}
\end{figure}

A combined study \cite{Thome:2004sk}, leaving both the primordial 
$k_{\perp}$ and the $\Lambda$ value free, still gives some preference 
to $\langle k_{\perp} \rangle = 2$~GeV and the standard 
$\Lambda^{(4)} = 0.192$~GeV, but differences relative 
to an alternative with $\langle k_{\perp} \rangle = 0.6$~GeV 
and $\Lambda^{(4)} = 0.22$~GeV are not particularly large, 
Fig.~\ref{resumandshower:figa}. 

\begin{figure}[pb]
\begin{center}
\vspace*{-5mm}
\includegraphics[width=12cm]{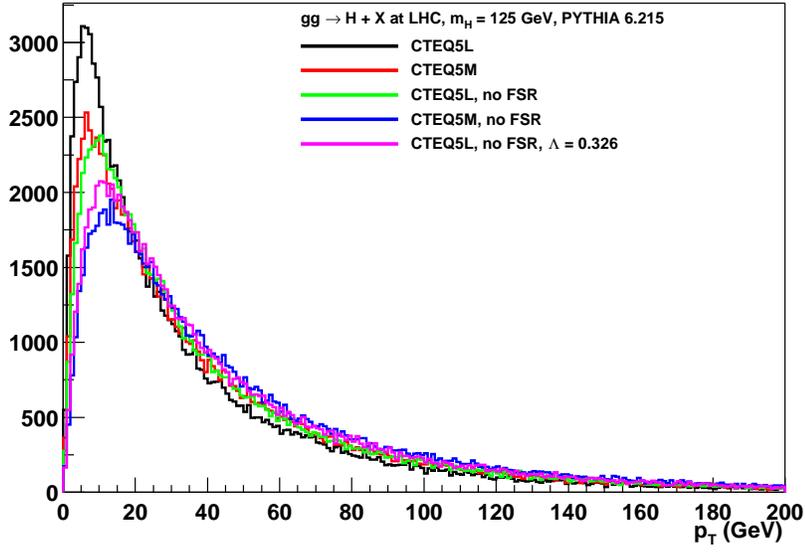}
\vspace*{-5mm}
\end{center}
\caption{Comparison of the \textsc{Pythia} $p_{\perp}$ distributions for 
Higgs production at the LHC using LO and NLO pdf's and turning timelike 
parton showering off (no FSR) and on.}  
\label{resumandshower:figb}
\end{figure}

\section{FURTHER COMPARISONS}

Returning to Higgs production at the LHC, in Fig.~\ref{resumandshower:figb} 
are shown a number of predictions for the current standard \textsc{Pythia} 
shower routines. Using CTEQ5M rather than CTEQ5L results in more  gluon 
radiation and a broader $p_{\perp}$ distribution due to the 
large value of $\Lambda$. Likewise turning off timelike showers for gluons
radiated from the initial state also results in the peak of the $p_{\perp}$ 
distribution moving outwards. 

\begin{figure}[htbp]
\begin{center}
\vspace*{-3mm}
\includegraphics[width=12cm]{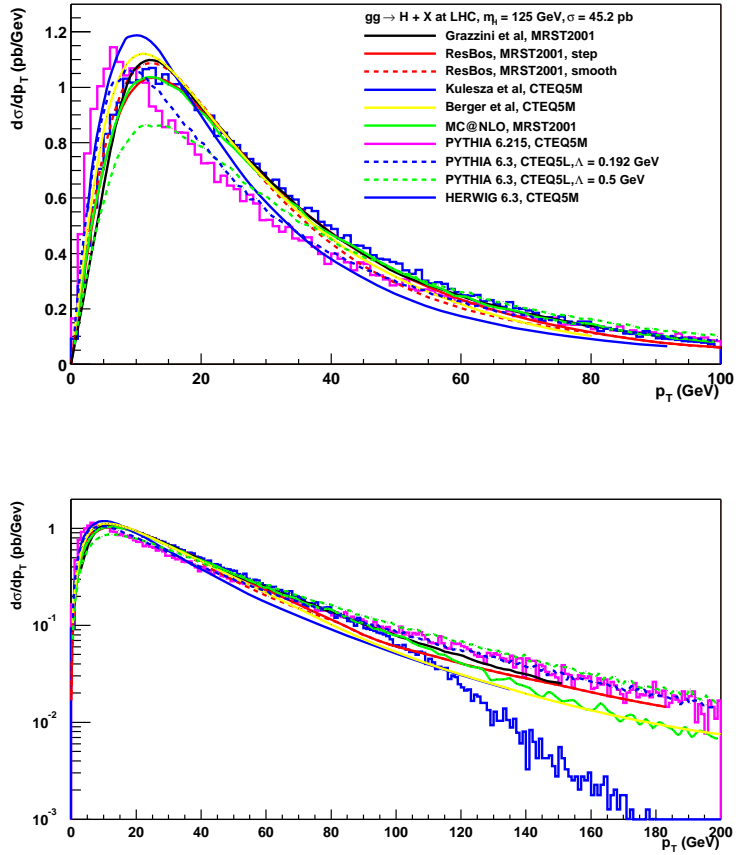}
\vspace*{-6mm}
\end{center}
\caption{Comparison of various $p_{\perp}$ distributions for Higgs 
production at the LHC. The curves denoted Grazzini \cite{Bozzi:2003jy}, 
ResBos \cite{Balazs:2000wv}, Kulesza \cite{Kulesza:2003wn} and Berger 
\cite{Berger:2002ut} are resummation descriptions, while 
MC\@NLO \cite{Frixione:2002ik,Frixione:2003ei},
\textsc{Herwig} \cite{Corcella:2000bw} and \textsc{Pythia} are 
generators, \textsc{Pythia} 6.3 referring to the new algorithm 
outlined above. }  
\label{resumandshower:figc}
\end{figure}

We can now compare the results with resummation descriptions and other
generators, Fig.~\ref{resumandshower:figc} \cite{higgshere}. As we see, 
the new \textsc{Pythia} routines agree  better with resummation 
descriptions than in the past \cite{Balazs:2000sz}, attesting to the 
importance of various minor technical details of the Monte Carlo approach. 
One must note, however, that some spread remains, and that it is not 
currently possible to give an unambiguous prediction.

\section{CONCLUSIONS}

We have studied $p_{\perp\mathrm{Z}}$ and $p_{\perp\mathrm{H}}$ spectra,
as a way of exploring perturbative and nonperturbative effects in hadronic
physics. Specifically, we have pointed out a number of ambiguities that
can exist in a shower approach, e.g.\ that the shower goes beyond the
simpleminded leading-log evolution and kinematics, while still making use 
of leading-log parton densities. Attempts to correct for mismatches in
general tend to increase the perturbative $p_{\perp\mathrm{Z}}$.
The need for an unseemly large primordial $k_{\perp}$ in the shower 
approach is thus reduced, but not eliminated. There is still room for,
possibly even a need of, perturbative evolution beyond standard DGLAP at 
small virtuality scales.  

\bibliography{biblifile}

\begin{thebibliography}{10}

\bibitem{Balazs:2000sz}
C.~Balazs, J.~Huston, and I.~Puljak.
\newblock {\em Phys. Rev.}, D63:014021, 2001.

\bibitem{Collins:1981uk}
J.C. Collins and D.E. Soper.
\newblock {\em Nucl. Phys.}, B193:381, 1981.

\bibitem{Sjostrand:2000wi}
T.~Sj{\"o}strand et~al.
\newblock {\em Comput. Phys. Commun.}, 135:238--259, 2001.

\bibitem{Sjostrand:2003wg}
T.~Sj{\"o}strand, L.~L{\"o}nnblad, S.~Mrenna, and P.~Skands.
\newblock 2003.
\newblock LU TP 03-38 [hep-ph/0308153].

\bibitem{Sjostrand:1985xi}
T.~Sj{\"o}strand.
\newblock {\em Phys. Lett.}, B157:321, 1985.

\bibitem{Bengtsson:1986gz}
M.~Bengtsson, T.~Sj{\"o}strand, and M.~van Zijl.
\newblock {\em Z. Phys.}, C32:67, 1986.

\bibitem{Miu:1998ju}
G.~Miu and T.~Sj{\"o}strand.
\newblock {\em Phys. Lett.}, B449:313--320, 1999.

\bibitem{Thome:2004sk}
E.~Thom{\'e}.
\newblock 2004.
\newblock LU TP 04-01 [hep-ph/0401121].

\bibitem{Lai:1999wy}
H.~L. Lai et~al.
\newblock {\em Eur. Phys. J.}, C12:375--392, 2000.

\bibitem{Sjostrand:2004qj}
T.~Sj{\"o}strand.
\newblock 2004.
\newblock These proceedings. LU TP 04-05 [hep-ph/0401061].

\bibitem{Affolder:1999jh}
T.~Affolder et~al.
\newblock {\em Phys. Rev. Lett.}, 84:845--850, 2000.

\bibitem{higgshere}
C.~Balazs et~al.
\newblock These proceedings.

\bibitem{Bozzi:2003jy}
G.~Bozzi, S.~Catani, D.~de~Florian, and M.~Grazzini.
\newblock {\em Phys. Lett.}, B564:65--72, 2003.

\bibitem{Balazs:2000wv}
C.~Balazs and C.~P. Yuan.
\newblock {\em Phys. Lett.}, B478:192--198, 2000.

\bibitem{Kulesza:2003wn}
A.~Kulesza, G.~Sterman, and W.~Vogelsang.
\newblock 2003.
\newblock [hep-ph/0309264].

\bibitem{Berger:2002ut}
E.L. Berger and J.~Qiu.
\newblock {\em Phys. Rev.}, D67:034026, 2003.

\bibitem{Frixione:2002ik}
S.~Frixione and B.R. Webber.
\newblock {\em JHEP}, 06:029, 2002.

\bibitem{Frixione:2003ei}
S.~Frixione, P.~Nason, and B.R. Webber.
\newblock {\em JHEP}, 08:007, 2003.

\bibitem{Corcella:2000bw}
G.~Corcella et~al.
\newblock {\em JHEP}, 01:010, 2001.

\end{thebibliography}

\end{document}